\documentclass[aps,pra,twocolumn,showpacs,superscriptaddress]{revtex4}
\usepackage{epsfig}
\usepackage{bm,amsmath,amssymb,latexsym,mathrsfs,graphicx,enumerate}
\usepackage[mathcal]{euscript}
\usepackage[pdftitle={Reaction-diffusion front crossing a local defect},pdfsubject={Reaction-diffusion front crossing a local defect},pdfkeywords={reaction diffusion},pdfauthor={Jean-Guy Caputo and Benoit Sarels},bookmarks=false,pdfcreator={dvipdf}]{hyperref}
\usepackage{psfrag}
\usepackage{float}

\newcommand{\be}{\begin{equation}}
\newcommand{\ee}{\end{equation}}
\begin{document}

\title{Reaction-diffusion front crossing a local defect}
\author{Jean-Guy Caputo}
\email{caputo@insa-rouen.fr}
\author{Benoit Sarels}
\email{benoit.sarels@math.cnrs.fr}
\affiliation{Laboratoire de Math\'ematiques \\
INSA de Rouen \\
Avenue de l'Universit\'e, B.P 08 \\
76801 Saint-\'Etienne du Rouvray, France}
\date{\today}

\begin{abstract}
The interaction of a Zeldovich reaction-diffusion front
with a localized defect is studied numerically and analytically.
For the analysis, we start from conservation laws and develop
simple collective variable ordinary differential equations for the
front position and width. Their solutions are in good agreement
with the solutions of the full problem. Finally using this
reduced model, we explain the pinning of the front on a large defect
and obtain a quantitative criterion.
\end{abstract}

\maketitle


\section{Introduction}

Since the pioneering work of Zeldovich et al \cite{zeldovich} reaction
diffusion equations have been important models to describe the combustion
of solid materials. The so-called combustion front connects two equilibria,
the unburnt solid on one side and the burnt solid on the other.
These models also describe in some limit the propagation of
a nerve impulse in a neuron. See \cite{scott} for a review of these
applications. Fisher\cite{fisher} in 1937 with his model of gene propagation
opened a new area of application. In all these systems there is one
component, a concentration which diffuses and a reaction term, which is
cubic for the Zeldovich model and quadratic for the Fisher equation.
From another point of view, these one variable systems can be seen
as a reduction of more general models like the Susceptible-Infected-Recovered
(SIR) model of Kermac-McKendrick \cite{km27} describing the propagation
of epidemics.  
An easy way to see this is to start from the Susceptible-Infected-Recovered (SIR) three-component quadratic model of Kermack--McKendrick \cite{km27}.
We assume $R=0$ so that $S = N-I$ where $N$ is the total number.
The quadratic terms are then $S~ I= (N-I)I$.
Therefore the Fisher or Zeldovich equations can be considered
as the simplest model of the propagation of an epidemic.

In many cases the reaction term is non-uniform in space. The powder
in a combustion tube can present some defects along the tube.
A sudden enlargement of an axon is known to block nerve impulse as
discussed in \cite{scott}. This effect was analyzed by Chapuisat et
al \cite{chapuisat}.
Geographical factors like rivers or mountains slow down an epidemic front. This
can be seen on the records of the plague epidemic in the 13th century.
See for example \cite{murray} and \cite{gaudart2009} on this example. These
factors also need to be considered when modeling the diffusion of
certain genes \cite{novembre}. Some defects are also time dependant
like seasonality effects.
In a number of situations, the defect can be considered as localized
in a given region of space. This is the case for a river in the example
of the propagation of an epidemic given above. Such a local defect
will enable to reduce the dimensionality of the problem. For example
a line defect in 2D will separate the plane in two so that one can
consider only the propagation along the 1D line normal to the defect line.
In the following article on the Zeldovich model
we study the interaction of a 1D front with a
localized defect. We consider a 1D model because
it is simple, one can do analysis and there is an exact front
solution if the reaction term is a third degree polynomial.
As mentioned above even if the problem is 2D or 3D, a localized
defect reduces the geometry to 1D: before the defect and after the defect.
This approximation makes sense of course for a narrow tube.
For this model, we show that the front can be
stopped by a large enough defect.
We introduce an approximate analysis
based on conservation laws which gives ordinary differential equations
for the front position and width. The solutions of these simple models
are in good agreement with the solution of the partial differential
equation. These collective variable equations easily lead to the solution
of the inverse problem of determining the defect from the motion of the
front.\\
The article is organized as follows. Section \ref{preliminary} presents a preliminary analysis
of the model. Numerical solutions are shown and analyzed in section \ref{numerical_analysis}.
Section \ref{collective_analysis} introduces the collective variable equations whose solutions
are compared to the solutions of the full equation in section \ref{comparison}. We conclude
in section \ref{conclusion}.

\section{\label{preliminary}Preliminary analysis}

A general reaction-diffusion equation in an heterogeneous medium is 
\be\label{rds}
u_t = u_{xx}+  s(x) R(u)
\ee
We consider that $R$ is a third degree polynomial with three real roots so that the model \eqref{rds} is written
\be\label{zeldo_s}
u_t = u_{xx}+  s(x) u(1-u)(u-a)
\ee
with $0<a<1$. The homogeneous states are $u^* \equiv 0,1,a$. Only the first two are stable\cite{scott} so we will consider fronts connecting 0 to 1. If $s$ is homogeneous, the front solution can be calculated by assuming that it is a traveling wave $u(z)\equiv u(x-ct)$ and assuming that $du/dz = Ku (1-u)$ where $K$ is a constant\cite{mornev,scott}.

One obtains the kink exact solution
\be\label{sol_exact}
u(x,t)=\frac{1}{1+\exp\left(\pm\sqrt{\frac{s}{2}}(x-ct)\right)} ,\ee
where the speed is
\be\label{kink_speed} c=\pm\sqrt{\frac{s}{2}}(1-2a).\ee
Depending on the $\pm$ sign we have a increasing (resp. decreasing) kink
going from $0$ (resp. $1$) as $x \to -\infty$ to $1$ (resp. $0$)
as $x \to +\infty$ for the $+$ (resp. $-$) sign.
The width of the kink is
\be\label{kink_width} w=\sqrt{\frac{2}{s}}.\ee
It is inversely proportional to the inhomogeneity so a large $s$
corresponds to a very sharp and fast front.

\section{\label{numerical_analysis}Numerical analysis for a localized defect $s(x)$}

When $s(x)$ varies there is only a local balance between the reaction term $R$ and the diffusion term $u_{xx}$. The equation does not have an explicit solution so we study it numerically. Specifically we use the method of lines where we discretize space by the finite difference method. The time advance is given by an ordinary differential equation solver. All the details are given in the appendix \ref{numerical_details}. For simplicity we chose the parameter $a=0.3$ throughout the
article.

The fronts are stable for many different values of the localized defect $s(x)$.
Precisely, the fronts continue to exist with a speed and a width that change.
Therefore it is reasonable to fit the solution at each time using a least square procedure by a kink
\be\label{k}
u_k(x,t)=\frac{1}{1+\exp\left(\frac{x-x_0(t)}{w(t)}\right)},
\ee
where the time dependence is only through the collective coordinates, the kink position $x_0$ and width $w$.
The details of the fitting procedure are given in the appendix \ref{numerical_details}.
The important fact is that the normalized least-square error is 
small ($ < 10^{-4}$ ) in all the numerical results. Therefore the fit
is very good.

We will consider two main localized defects : a gaussian defect and a tanh defect.
There are two length scales in the problem, the kink width $w$ given by
\eqref{kink_width} in the homogeneous case and the characteristic 
length $d$ of the defect. For a given front, the defect will be wide or narrow
depending on the ratio of these two length scales. 

\subsection{Gaussian defect}
\begin{figure}[H]
\psfrag{x0}[r][r]{$x_0$}
\centerline{
\epsfig{file=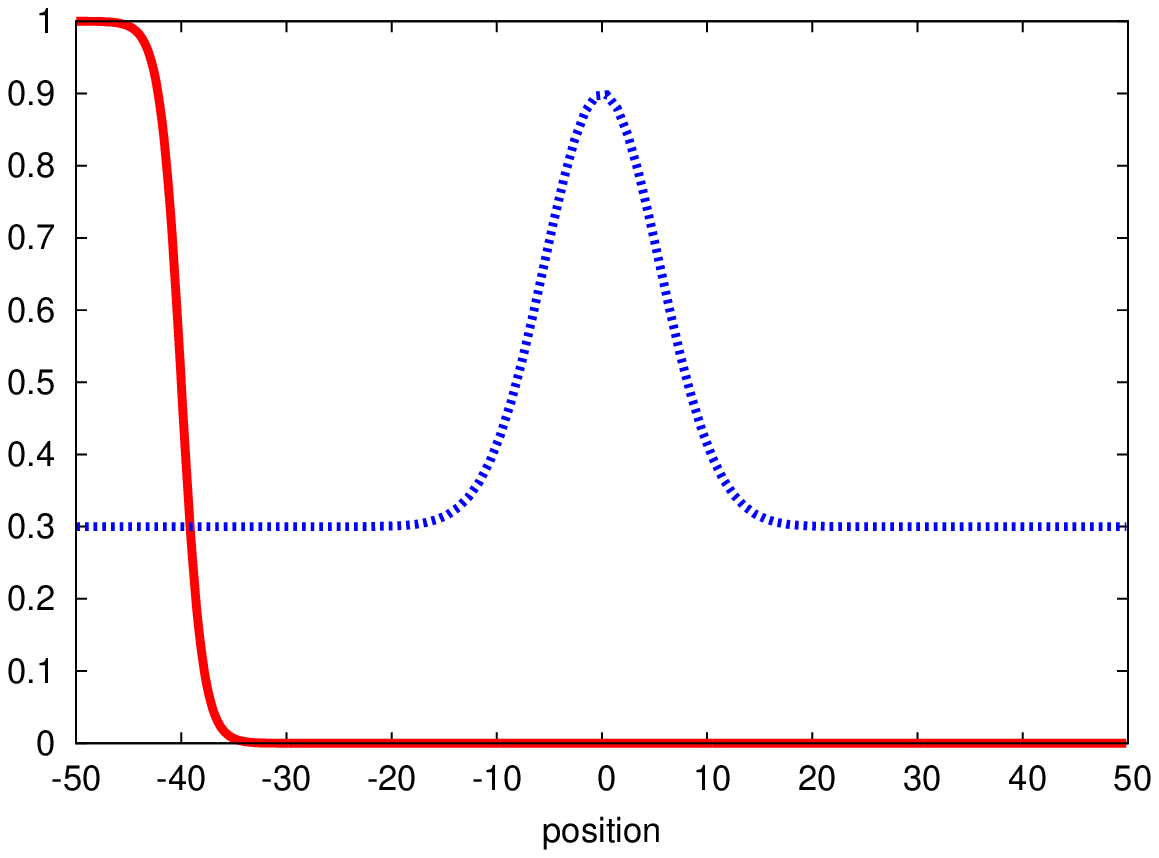,width=0.5\linewidth,angle=0}
\epsfig{file=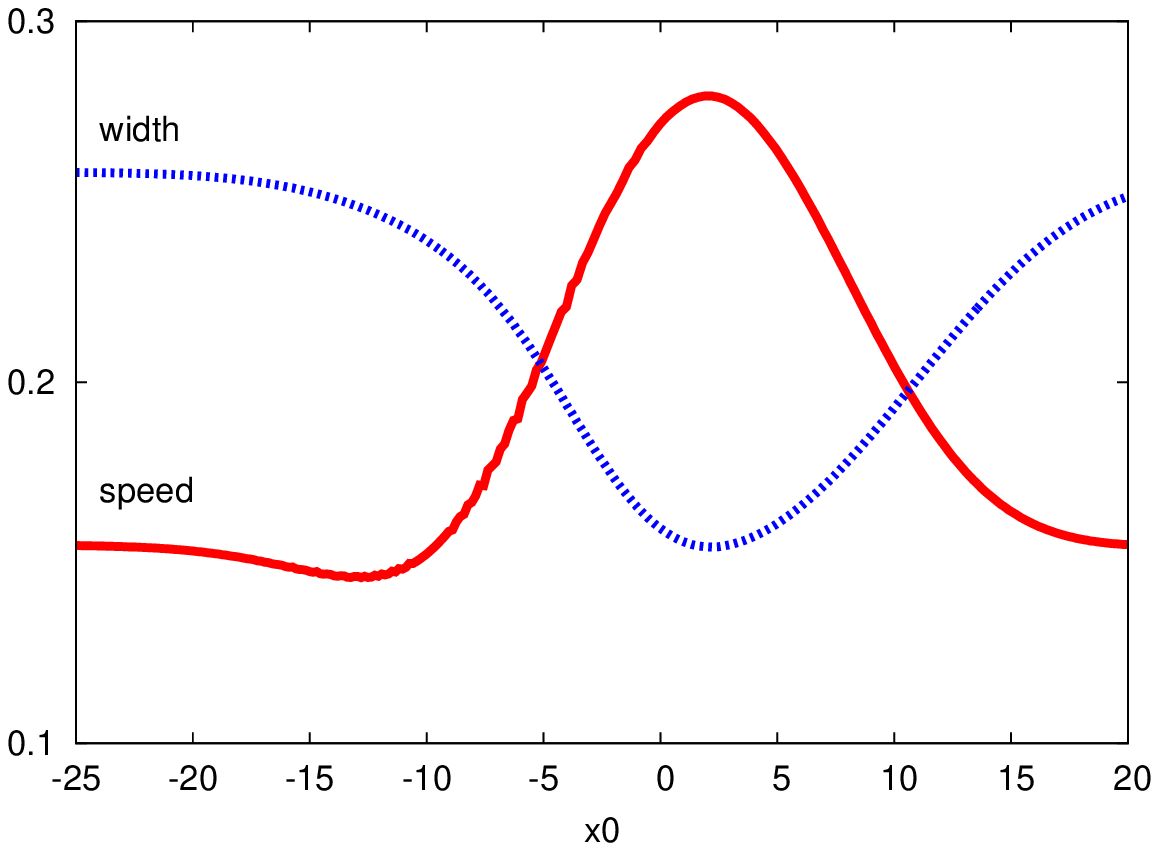,width=0.5\linewidth,angle=0}
}
\caption{Adiabatic motion of a front.
The left panel shows the initial front $u(x,t=0)$ together with the defect $s(x)$.
The right panel shows the speed $x_0'$ and width $w$ of the kink as a function of the kink position $x_0$.
The parameters are $s_0=0,3$, $s_1=0,6$ and $d = 30$. 
}
\label{f1}
\end{figure}

Specifically the first class of defect is given by
\be\label{gaus_defect}
s(x) = s_0 + s_1 \: \exp \left({\frac{-x^2}{2d}} \right)
\ee
When the defect varies on a length scale large compared with the width of the kink, the front moves adiabatically.
An example is shown in figure \ref{f1}.
The left panel shows the initial front together with the gaussian defect as a function of position $x$.
The right panel shows the speed $x_0'$ and width $w$ of the kink as a function of the kink position $x_0$.
The defect $s(x_0)$ is also reported in this graph.
The speed is computed using a centered difference approximation from the time-series $x_0(t)$.
One sees that the kink accelerates and gets steeper as it runs into the defect. 

\begin{figure}[H]
\psfrag{x0}[r][r]{$x_0$}
\centerline{
\epsfig{file=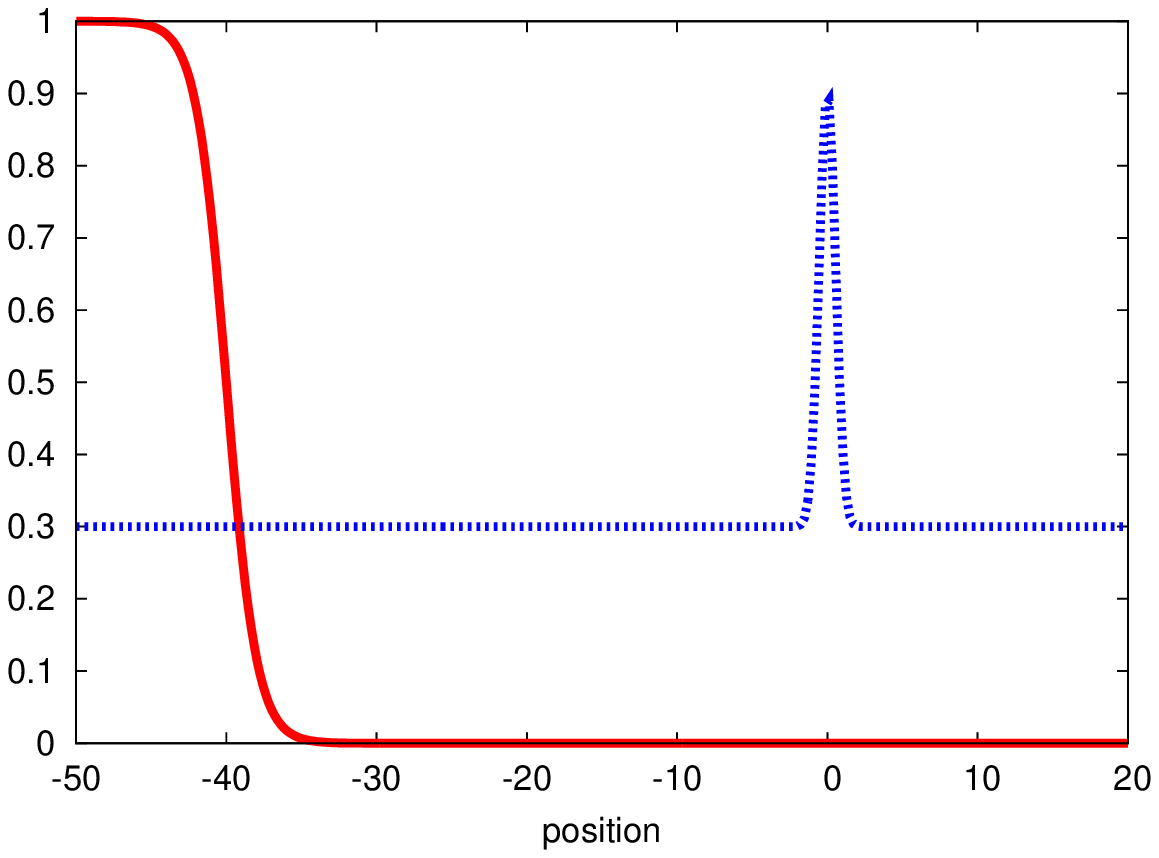,width=0.5\linewidth,angle=0}
\epsfig{file=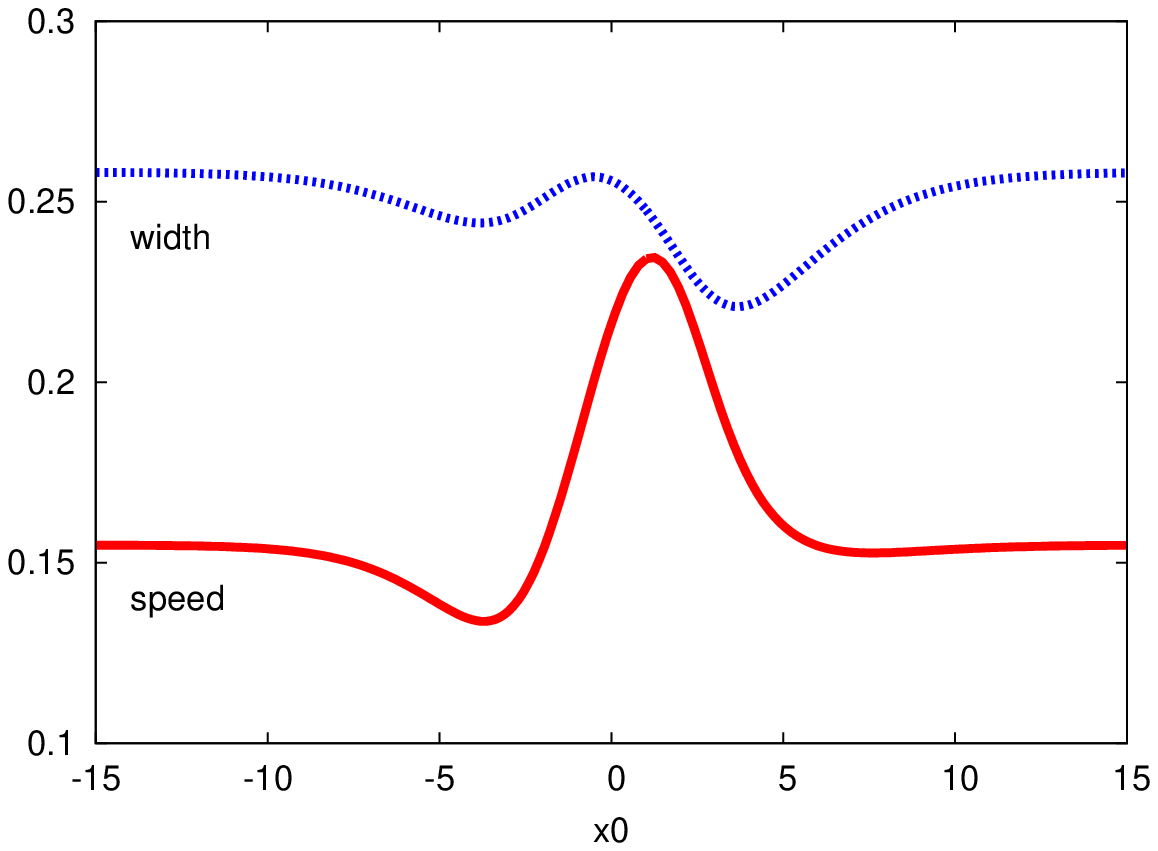,width=0.5\linewidth,angle=0}
}
\caption{Motion of a front as it hits a narrow defect.
The left panel shows the initial front $u(x,t=0)$ together with the defect $s(x)$.
The right panel shows the speed $x_0'$ and width $w$ of the kink as a function of the kink position $x_0$.
The parameters are $s_0=0,3$, $s_1=0,6$ and $d = 0,3$.
}
\label{f2}
\end{figure}

Now let us consider the other limiting case, i.e. when the defect is narrower than the kink as shown in the left panel of figure \ref{f2}. 
Here again the kink is able to breach the obstacle although it slows down before it and accelerates past it as shown in the right panel of figure \ref{f2}.
Notice also the modulation of the width.

\begin{figure}[H]
\psfrag{x0}[r][r]{$x_0$}
\centerline{
\epsfig{file=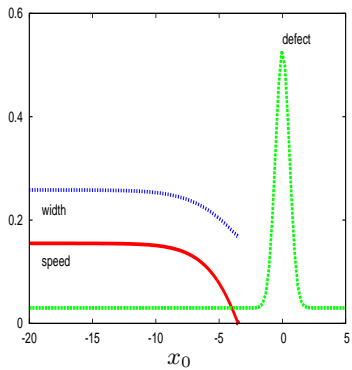,height=5cm,width=8cm,angle=0}
}
\caption{Pinning of the front when it hits a large defect.
The parameters are $s_0=0,3$, $s_1=7$ and $d = 0,3 $.}
\label{f3}
\end{figure}

If the amplitude of the defect is larger, the kink can be stopped as shown in figure \ref{f3}.
The left panel presents a defect with an amplitude $s_1=7$.
The right panel shows the width and the speed of the front.
The latter goes to zero and the width remains stationary.
To show that this effect is intrinsic we varied systematically the spatial resolution of the computation from $N=800$ to $N=12000$.
For all these runs the front stopped at the same position $x_0\approx-2,73$, with a width $w\approx1,80$.
This phenomenon is also called "pinning of the front" \cite{scott}.
We will analyze it in detail below.

\subsection{Tanh defect}
\begin{figure}[H]
\psfrag{x0}[r][r]{$x_0$}
\centerline{
\epsfig{file=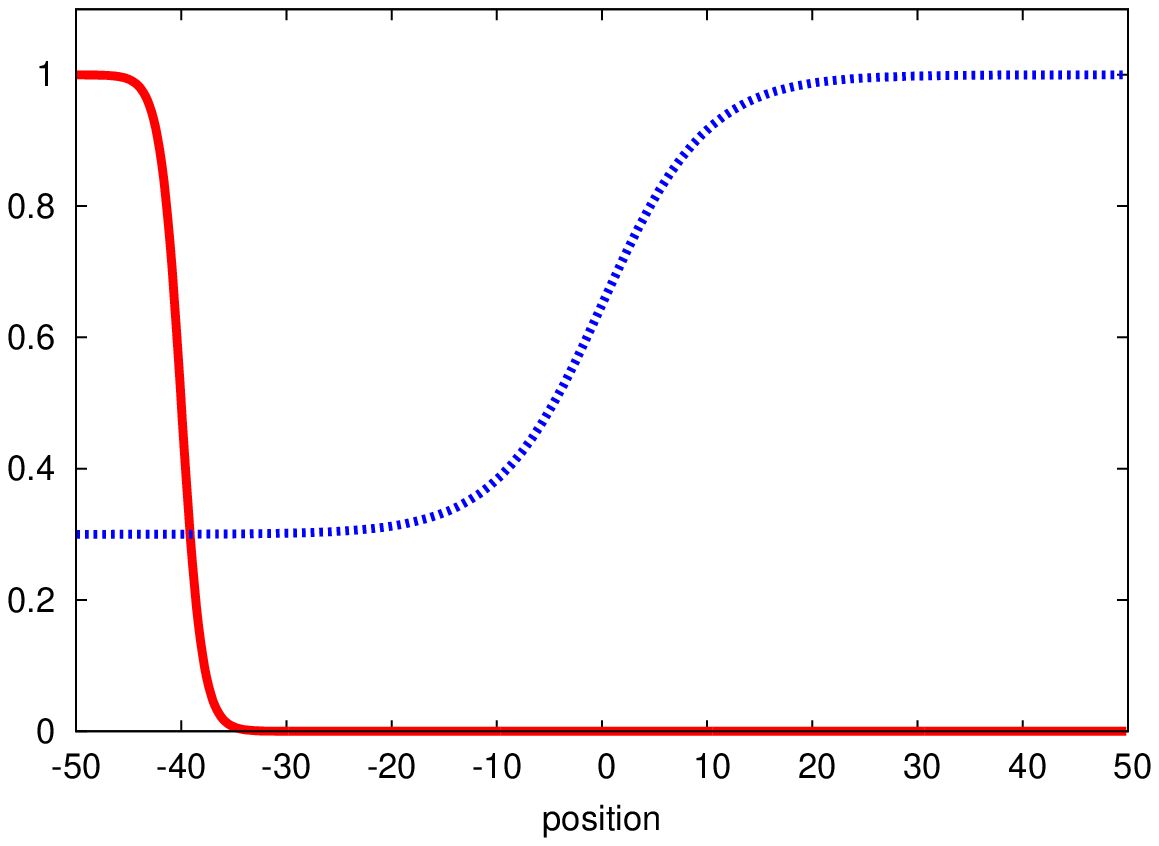,width=0.5\linewidth,angle=0}
\epsfig{file=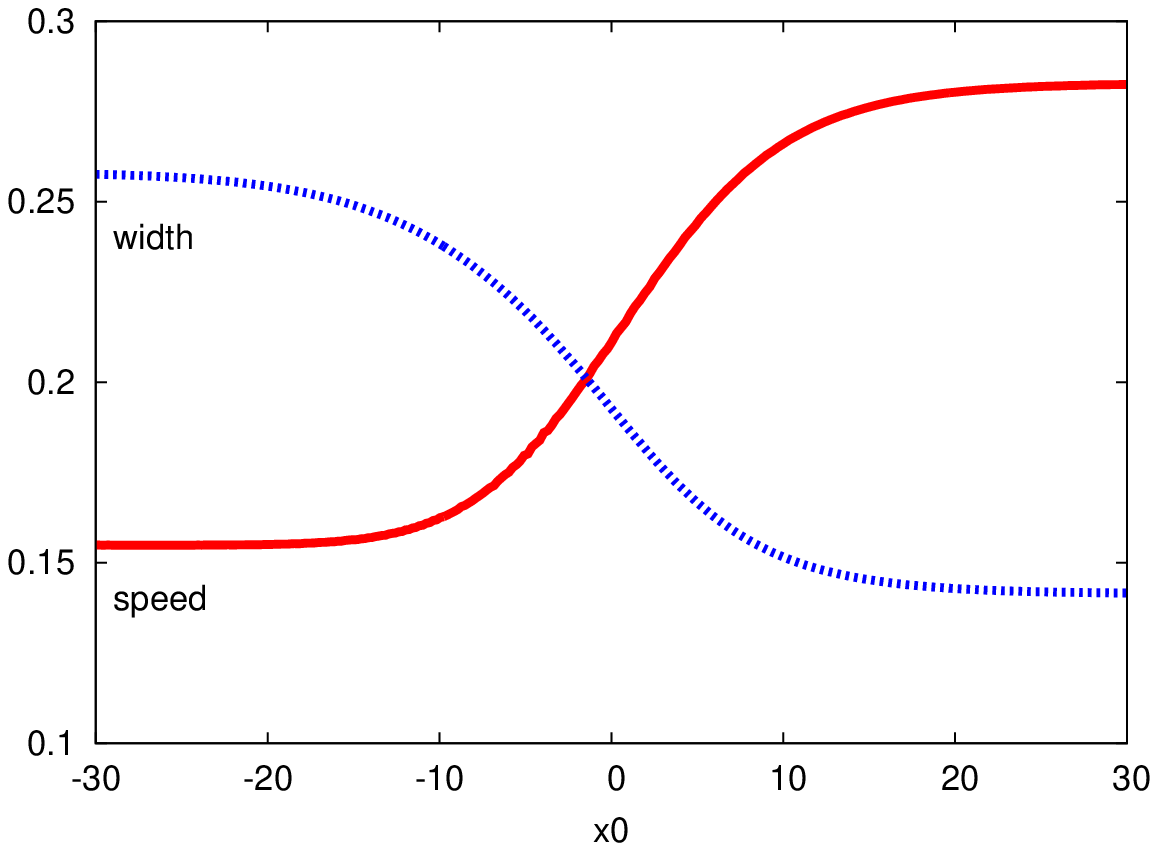,width=0.5\linewidth,angle=0}
}
\caption{Interaction of the front with a wide "tanh" defect. The parameters are $s_l=0,3$, $s_r=1$ (jump of $0,7$) and $d=10$.}
\label{f4}
\end{figure}

\begin{figure}[H]
\psfrag{x0}[r][r]{$x_0$}
\centerline{
\epsfig{file=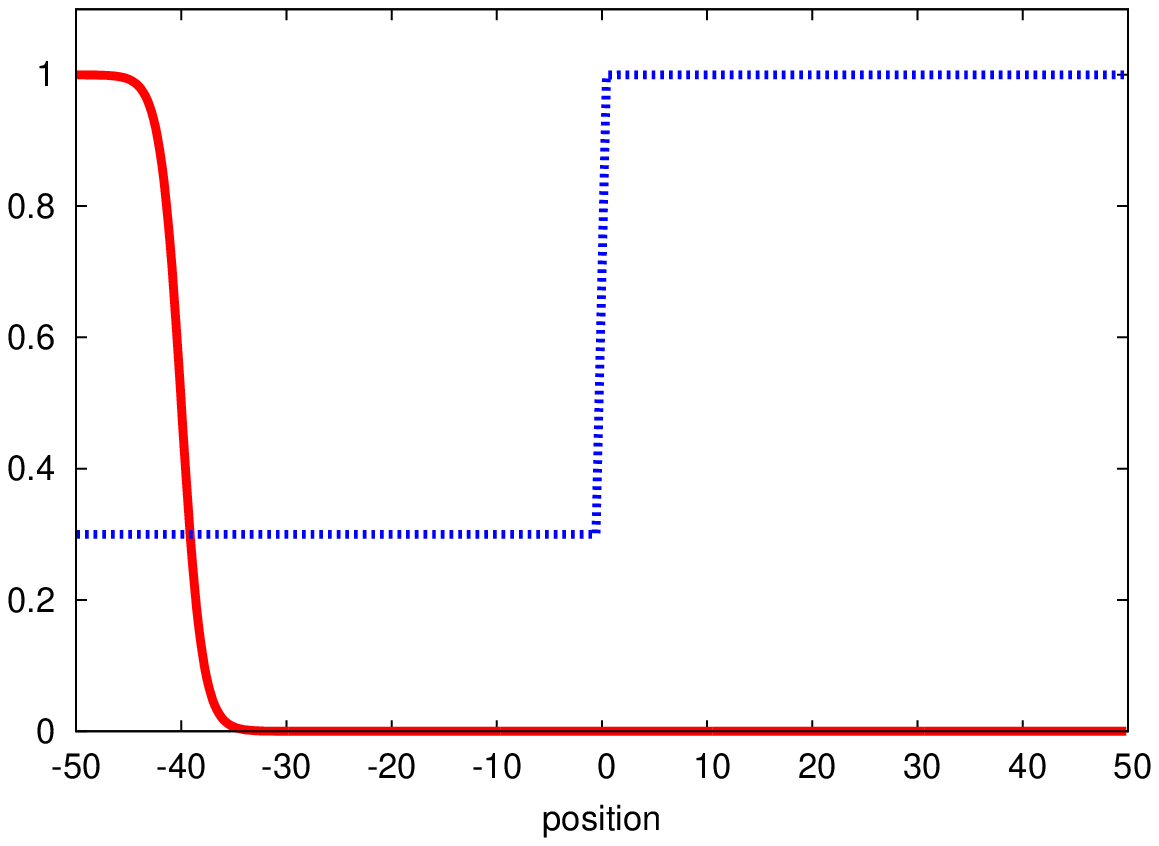,width=0.5\linewidth,angle=0}
\epsfig{file=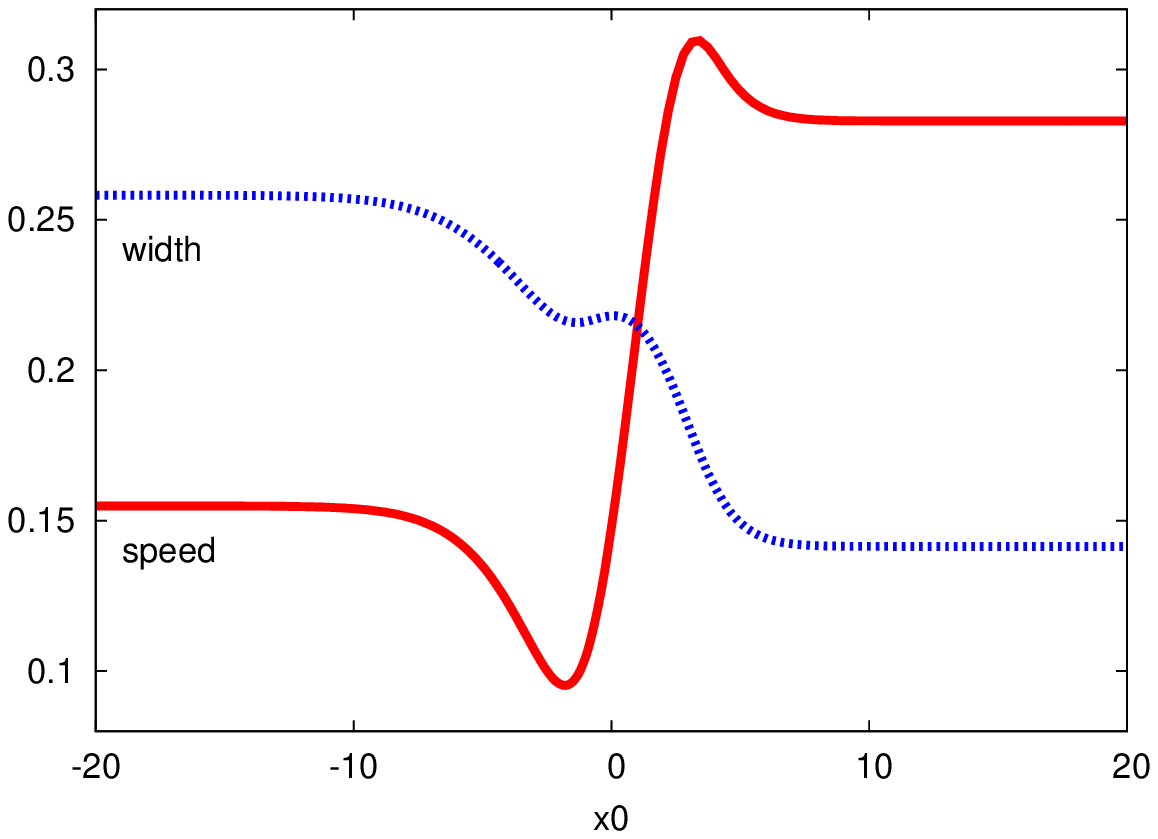,width=0.5\linewidth,angle=0}
}
\caption{Interaction of the front with a narrow "tanh" defect. The parameters are $s_l=0,3$, $s_r=1$ (jump of $0,7$) and $d=0,1$.}
\label{f5}
\end{figure}

We now consider a different type of defect which connects two different values $s_l$ on the left and $s_r$ on the right.
\be\label{tanh_def} 
s(x) = s_l + \frac{s_r-s_l}{2} \left(1+\tanh{\frac{x}{d}} \right)
\ee
We observe similar results as for the gaussian defect.
For a "wide" defect the front moves adiabatically as shown in figure \ref{f4}.
It speeds up and its width decreases as $s$ increases from left to right.
When the defect is narrow, the speed and width become non monotonic.
The results are shown in figure \ref{f5}.
Again the results have been checked by varying systematically the spatial resolution.
As previously the dip in the velocity indicates that increasing 
the amplitude we will obtain the pinning of the front. This indeed happens.
We will show in section \ref{critere} that we can infer a simple criterion for the pinning of the front.

\section{\label{collective_analysis}Collective coordinate analysis}

The numerical results in the previous section can be understood in a simple way.
Assume that the solution is a function of the reduced variable $z=x-ct$ where the speed $c$ is a slowly varying function of time.
Then reporting $u$ into the partial differential equation \eqref{rds} and integrating we get
\begin{multline}
-c \int_{-\infty}^{+\infty} U'(x-ct) dx
\\
=
\int_{-\infty}^{+\infty} U''(x-ct) dx 
+ \int_{-\infty}^{+\infty} s(x)R(u(x,t)) dx
\end{multline}
Consider a front going from 1 (to the left) to 0. Since the 
front is flat away from the defect we have
$$\int U''(z) dz = 0,~~\int U'(z) dz = -1,$$
so that we obtain
\begin{equation} \label{c1}
c = \int_{-\infty}^{+\infty} s(x)R(u(x,t)) dx .
\end{equation}
This result shows that the speed decreases (resp. increases) 
when the front reaches a region where $s$ is smaller (resp. larger).
This is not completely correct because the width of the front also
depends on $s$ as shown in \eqref{kink_width}.

A more general approach is then to assume that the front keeps its
general form but that its center $x_0$ and width $w$ become modulated
\be\label{front_ansatz}
u(x,t) = U\left(\frac{x-x_0(t)}{w(t)}\right)\equiv U(z).
\ee
At this point we do not specify the form of $U(z)$.
This method is often called "variational approach" 
in the context of Hamiltonian partial differential equations \cite{scott}
which are derived from a Lagrangian density.
Here however there is no variational principle so we need to generate
two conservation laws to obtain the evolution of $x_0$ and $w$.
The first one is the original partial differential equation \eqref{rds}.
The second conservation law can be obtained by multiplying the equation
by $u$ 
\be\label{urds}
uu_t = u u_{xx} + s(x) u R(u).
\ee
In principle we could have also multiplied by $u_x$ but this would give
the wrong evolution if $s$ is singular. 
After replacing the expression \eqref{front_ansatz} into the two
partial differential equations \eqref{rds},\eqref{urds}, we get
the evolutions of the front position $x_0$ and width $w$
\begin{multline} \label{gen_dx0dw}
x_0' \int U' + w' \int U' z + w \int s(wz+x_0) R = 0,
\\
x_0' \int U U' + w' \int U U' z + w \int s(wz+x_0) U R = \frac{1}{w}\int U'^2,
\end{multline}
where the terms $\int U',~\int U' z, \int U U',~\int U U' z$ are 
integrals with respect to the variable $z$ and therefore are numbers.
The integrals
$$\int s(wz+x_0) R \equiv \int_{\infty}^{+\infty} s(wz+x_0)R(U(z))dz,$$
$$\int s(wz+x_0) U R\equiv \int_{\infty}^{+\infty} s(wz+x_0)U(z) R(U(z))dz,$$
depend on $x_0$ and $w$.
They yield the source terms for the differential equations.
All the details are given in Appendix B.
These ordinary differential equations are very general and can be transposed to different types of nonlinearities.
The only assumptions are that the defect $s(x)$ is localized and that the front keeps its functional profile.

We consider the simplest ansatz \eqref{k} which is suitable for the Zeldovich equation because it is an exact solution when $s$ is constant.
Then \eqref{gen_dx0dw} reduce to
\begin{equation}
\left\{
\begin{aligned}
\label{xtwt_gen}
x_0' &= w \int_{-\infty}^{+\infty} s(wz+x_0)R(U(z)) dz ,\\
w'   &= \frac{1}{3w} + w  \int_{-\infty}^{+\infty} s(wz+x_0) (1-2U(z))R(U(z)) dz .
\end{aligned} 
\right.
\end{equation}
Note that we still have made no assumptions on the nonlinearity $R$.

We will consider three main situations, a defect that varies on a scale longer than the width of the front and two sharp defects represented respectively as a Dirac distribution and a Heaviside function.
For a defect that varies on a scale longer than $w$, $s(x_0+wz)\approx s(x_0)$ so that $s$ goes out of the integral, modifying the equations subsequently 
\begin{equation} \label{ode_adia}
\left\{
\begin{aligned}
x_0' &= \left( \frac{1-2a}{2} \right) w s(x_0), \\
w'   &= \frac{1}{3w} - \frac{w}{6} s(x_0).
\end{aligned} 
\right.
\end{equation}
These equations justify the notion of a local speed and width of the front.
When $s(x)=s$ is constant, we recover the results \eqref{kink_speed}, \eqref{kink_width}. Another remark is that here the defect can be extended. The only
condition that we require is that the front has reached it's equilibrium
state before reaching the defect region. \\
Note that the approximation can be sharpened by use of a Taylor expansion : 
$s(x_0+wz)\approx s(x_0)+w z s'(x_0)$.
Then the equations take the form
\begin{equation}
\left\{
\begin{aligned}
x_0' &= \frac{(1-2a)w}{2} s(x_0) - \frac{w^2}{2} s'(x_0), \\
w'   &= \frac{1}{3w} - \frac{w}{6} s(x_0) + \frac{(1-2a)w^2}{2}s'(x_0).
\end{aligned} 
\right.
\end{equation}

\begin{figure}[H]
\centerline{\epsfig{file=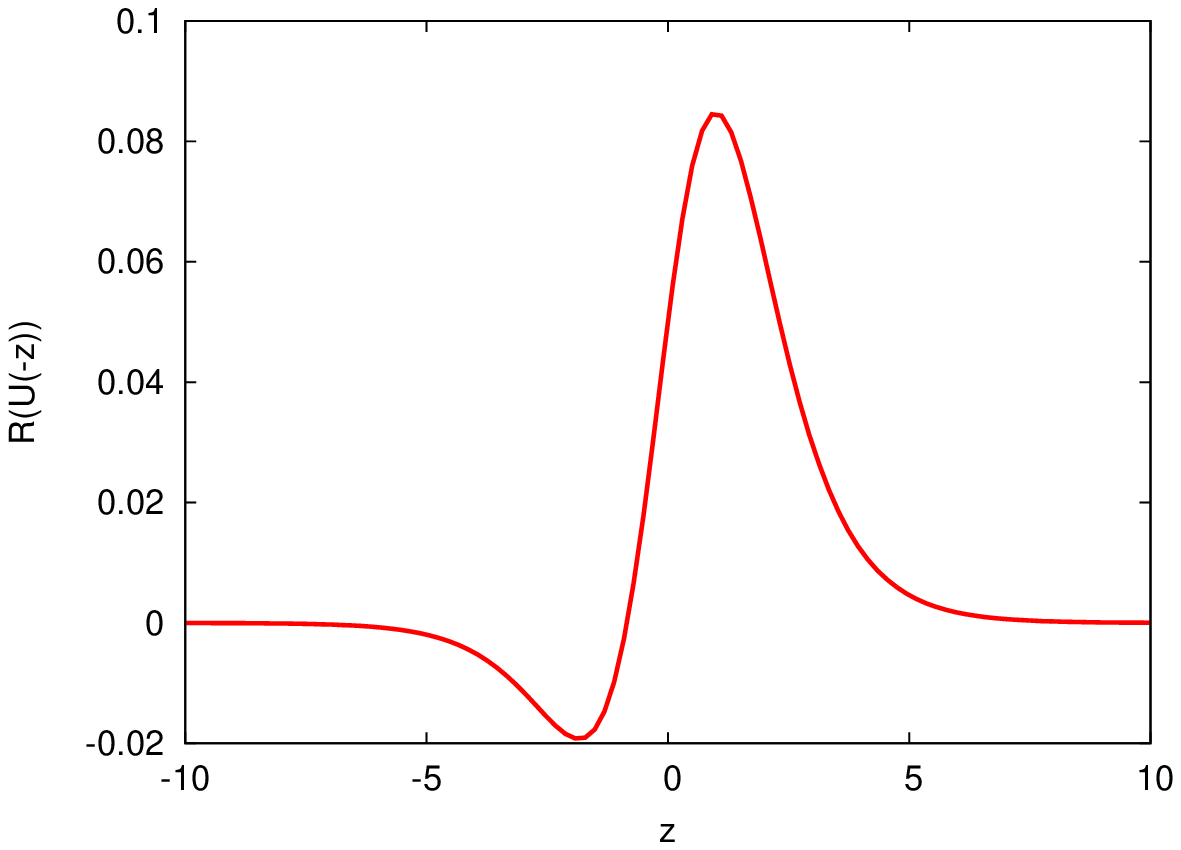,width=0.5\linewidth,angle=0}
\epsfig{file=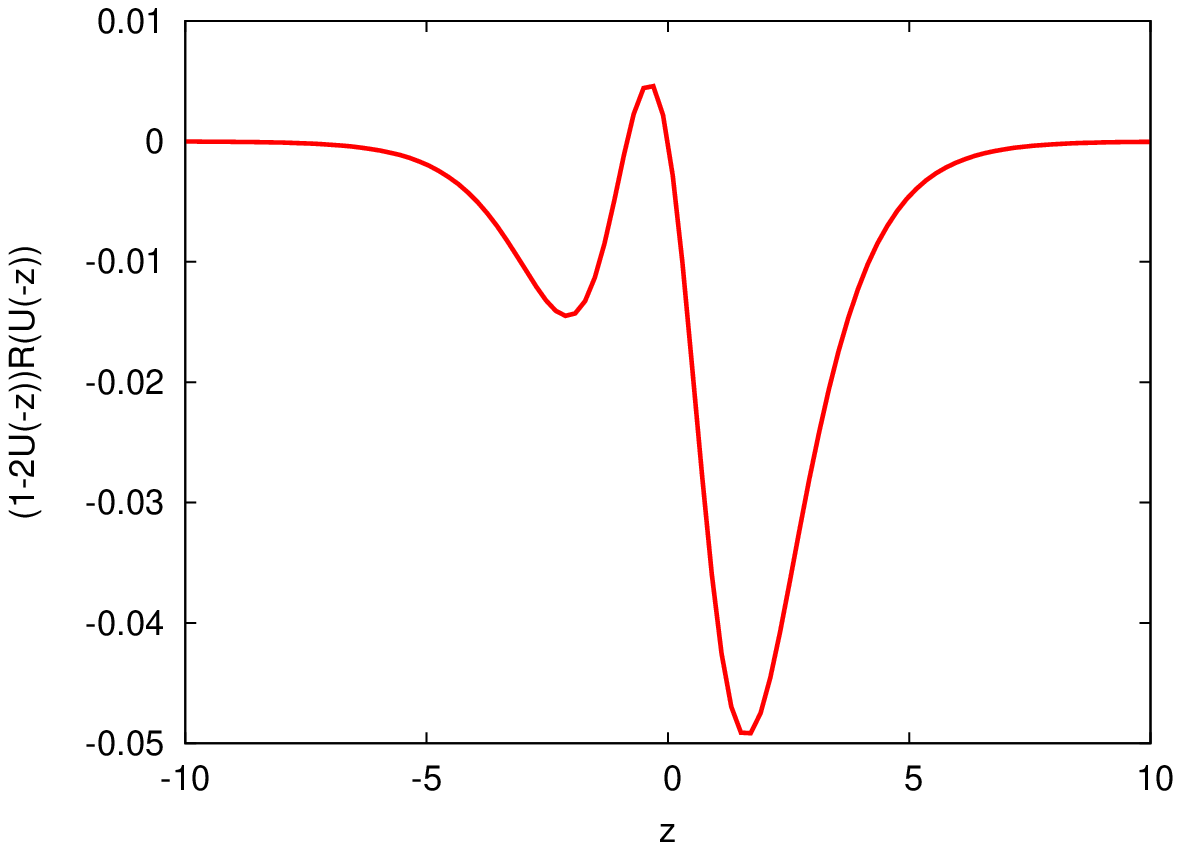,width=0.5\linewidth,angle=0}
}
\caption{Plot of the source terms $R(U(-z))$ (left panel) and 
$(1-2U(-z))~R(U(-z))$ (right panel) for a Dirac distribution defect 
as a function of the reduced variable $z$, for a chosen parameter $a=0.3$.}
\label{f_source_dirac}
\end{figure}

We now assume a sharp "bump-like" defect described by 
$s(x)=\alpha+\beta\delta(x)$ where $\delta(x)$ is the Dirac distribution
centered at $x=0$.
The system of equations is now reduced to 
\begin{equation} \label{ode_dirac}
\left\{
\begin{aligned}
x_0' &= \alpha w \left(\frac{1-2a}{2} \right) + \beta R\left(U\left(\frac{-x_0}{w}\right)\right) , \\
w'   &= \frac{1}{3w} -  \alpha \frac{w}{6}  + \beta \left(1-2 U\left(\frac{-x_0}{w}\right)\right) R\left(U\left(\frac{-x_0}{w}\right)\right) .
\end{aligned} 
\right.
\end{equation}
The source terms are given by 
\begin{equation}
\left\{
\begin{aligned}
R(U(z)) &= -e^z \frac{-1+a+a e^z}{(1+e^z)^3},\\
(1-2U(z))~R(U(z)) &= e^z (1-e^z) \frac{-1+a+a e^z}{(1+e^z)^4},
\end{aligned} 
\right.
\end{equation}
They are shown in Fig. \ref{f_source_dirac}.
Note how the two sources terms $R(U(-z))$ and $(1-2U(-z))~R(U(-z))$ are 
not symmetric.
Because $R(U(-z))$ is largely positive a defect will cause an acceleration 
of the kink for $\beta >0$.
But because there is also a negative part, the pinning of the front 
becomes possible both for $\beta >0$ and $\beta  <0$, see 
section \ref{critere} for details.
Notice also how $(1-2U(-z))~R(U(-z))$ is largely negative so that the 
width of the front decreases as it hits the defect for $\beta >0$.

\begin{figure}[H]
\centerline{
\epsfig{file=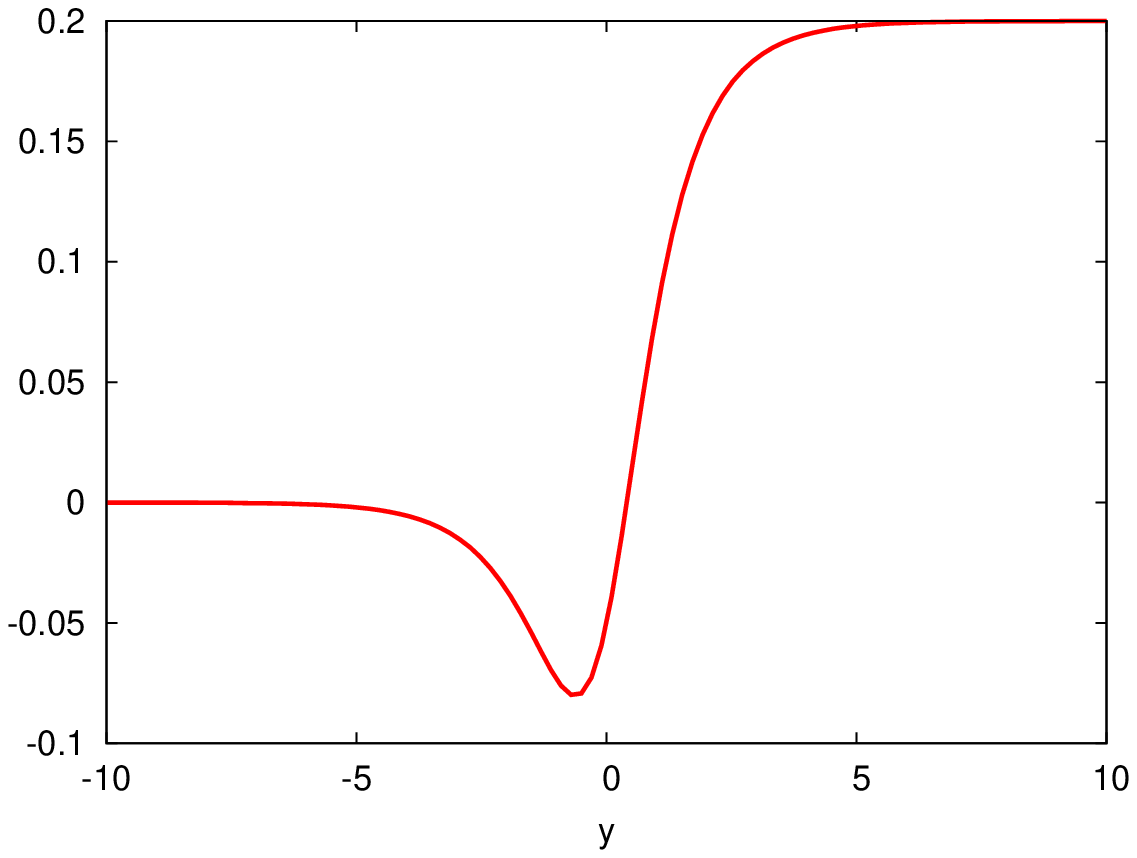,width=0.5\linewidth,angle=0}
\epsfig{file=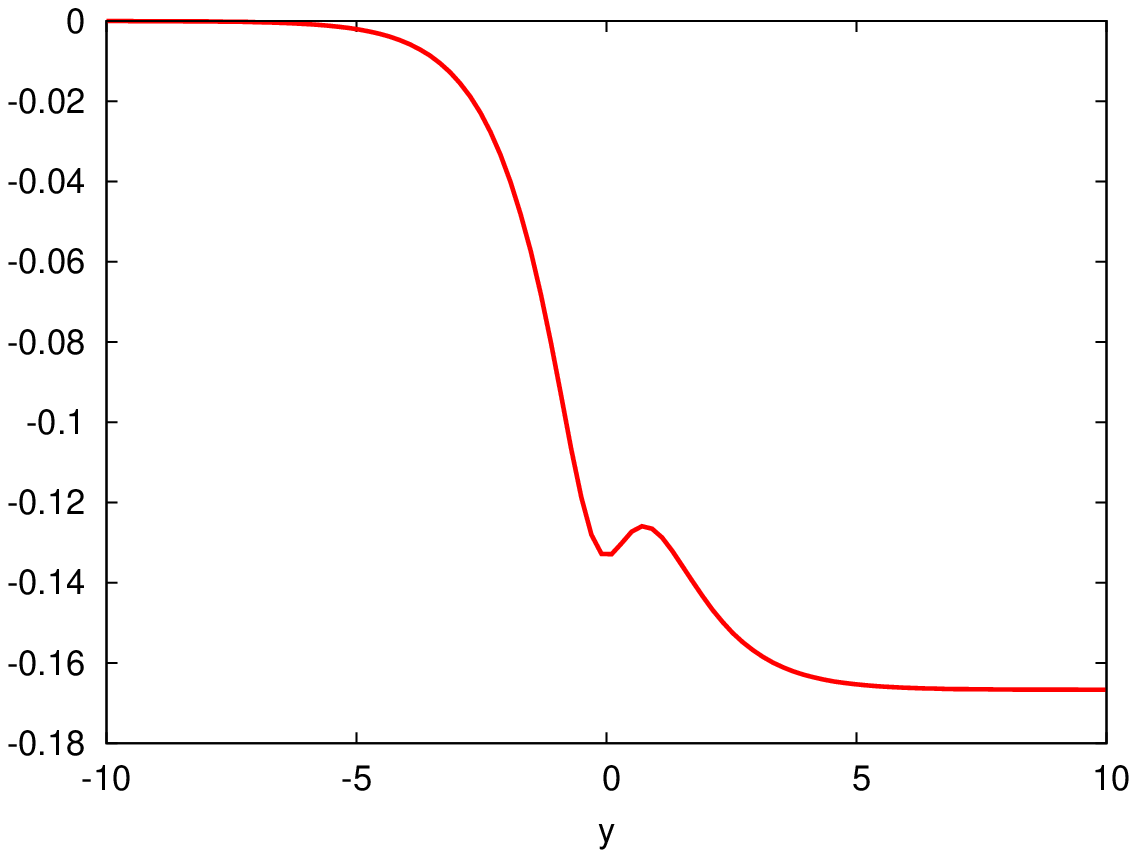,width=0.5\linewidth,angle=0}
}
\caption{Plot of the source terms $\int_{-y}^{+\infty} R(U(z))dz$ (left panel)
and $\int_{-y}^{+\infty} (1-2U(z))~R(U(z))dz$ (right panel) for a Heaviside distribution defect as a function of the reduced variable $y $, for a chosen parameter $a=0.3$.}
\label{f_source_heaviside}
\end{figure}

The other sharp defect that we study is $s(x)=\alpha+\beta H(x)$ where $H(x)$ is the Heaviside function.
The system of equations is now reduced to 
\begin{equation}\label{ode_heaviside}
\left\{
\begin{aligned}
x_0' &= \alpha w \left(\frac{1-2a}{2} \right) + \beta w \int_{\frac{-x_0}{w}}^{+\infty} R(U(z))\,dz ,\\
w'   &= \frac{1}{3w} -  \alpha \frac{w}{6}    + \beta w \int_{\frac{-x_0}{w}}^{+\infty} (1-2 U(z)) R(U(z))\,dz.
\end{aligned} 
\right.
\end{equation}
Here the source terms are
\begin{equation}
\left\{
\begin{aligned}
\int_{y}^{+\infty} R(U(z))dz           = {} & \frac{1}{2(1+e^y)^2}- \frac{a}{1+e^y}, \\
\int_{y}^{+\infty} (1-2U(z))~R(U(z))dz = {} & -\frac{2}{3(1+e^y)^3} +\frac{1+2 a}{2(1+e^y)^2} \\
                                            & -\frac{a}{1+e^y} .
\end{aligned}
\right.
\end{equation}
They are plotted in Fig. \ref{f_source_heaviside}.
The general remarks on the acceleration and width reduction of the front still apply.

To summarize, we have obtained fairly simple ordinary differential equations for the evolution of the front position $x_0$ and width $w$ for the Zeldovich equation.
We will see in the next section that these equations yield very good approximations to the solution of the partial differential equation.

\section{\label{comparison}Comparison between the full model and the reduced model}

To establish the validity of the reduced model, it is important to compare its solutions to the ones of the partial differential equation.
As discussed in the previous section, we classify the defects $s(x)$ as wide or narrow depending whether $w/d \ll 1 $ or $w/d \gg 1 $ where $w$ is the initial width of the front and $d$ is the width of the defect as defined in formulas \eqref{gaus_defect} and \eqref{tanh_def}.

\subsection{Adiabatic case}

\begin{figure}[H] 
\psfrag{x0}[r][r]{$x_0$}
\centerline{
\epsfig{file=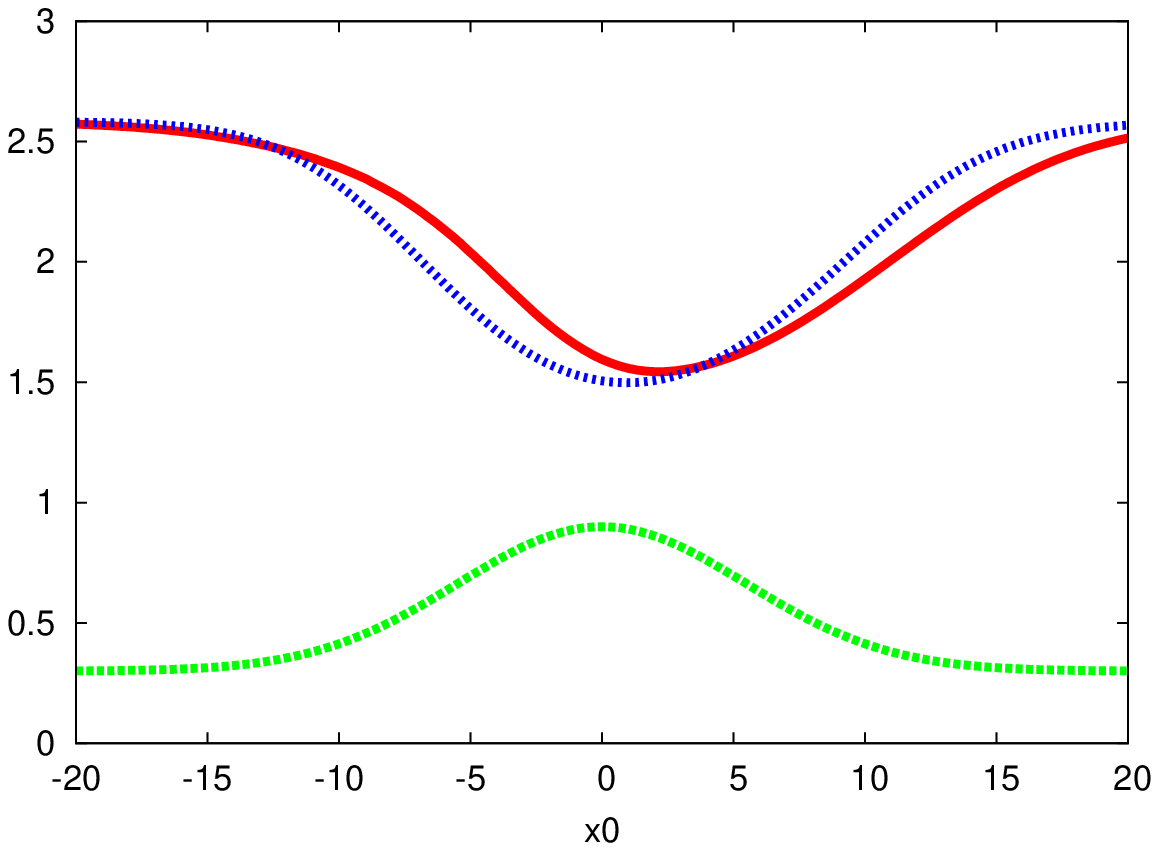,width=0.5\linewidth,angle=0}
\epsfig{file=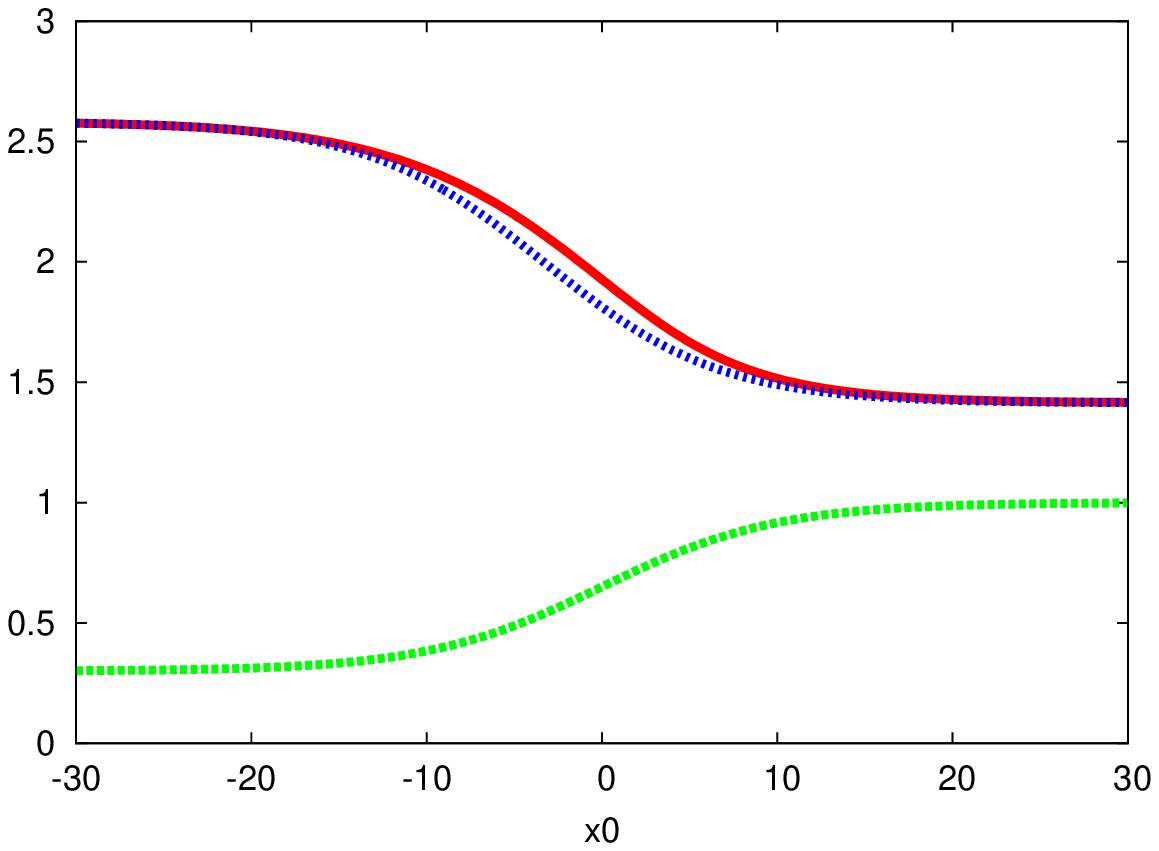,width=0.5\linewidth,angle=0}
}
\caption{Plots $(x_0,w)$ for the partial differential
equation \eqref{zeldo_s} in continuous line (red online) 
and the simple collective variable equations \eqref{ode_adia}
in dashed line (blue online). The left panel (resp. right panel) 
corresponds to a gaussian (resp. tanh) defect. The defects are
shown at the bottom of each panel.}
\label{f8}
\end{figure}

\begin{figure}[H] 
\psfrag{x0}[r][r]{$x_0$}
\centerline{
\epsfig{file=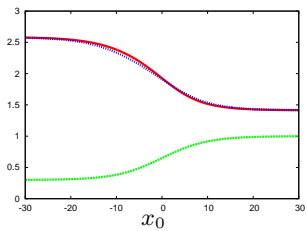,height=5cm,width=8cm,angle=0}
}
\caption{Plot $(x_0,w)$ for the partial differential
equation \eqref{zeldo_s} in continuous line (red online)
and the collective variable equations \eqref{gen_dx0dw}
in dashed line (blue online) for the "tanh" defect.}
\label{f9}
\end{figure}

When the defect is wide, equations \eqref{ode_adia} give evolutions
of the front position $x_0$ and width $w$ that are close
to the fits obtained from the solutions of the partial differential
equation . The plots are shown in
Fig. \ref{f8} for the gaussian defect of Fig. \ref{f1}
and the tanh defect of Fig. \ref{f4}.
The collective variable estimates can be improved by
calculating numerically the integrals in equation \eqref{gen_dx0dw}.
We compare in Fig. \ref{f9} the results for the
partial differential equation \eqref{zeldo_s} and for the
solutions of \eqref{gen_dx0dw} for the "tanh" defect. As can be seen the
agreement is excellent.

\subsection{Narrow defect: pinning }

We now consider that the defect width $d$ is smaller than the
front width $w$. Then the collective variable equations can
be reduced as shown in the previous section. 
We approximate the gaussian 
$s(x) = s_0 + s_1 \exp \left({\frac{-x^2}{2 d}} \right)$ by a 
Dirac delta function $s(x)=\alpha+\beta\delta(x)$. 
We choose $\alpha=s_0$ and $\beta=s_1\sqrt{2\pi d}$ so that 
$$ \beta = \int_{-\infty}^{+\infty} s_1 \exp \left({\frac{-x^2}{2 d}} \right)dx$$
The integrals associated to the defect are then equal. We tested the validity
of this approximation and found that the solutions agree to about
1 \% when $w > 10 d$.

\begin{figure}[H] 
\psfrag{x0}[r][r]{$x_0$}
\centerline{
\epsfig{file=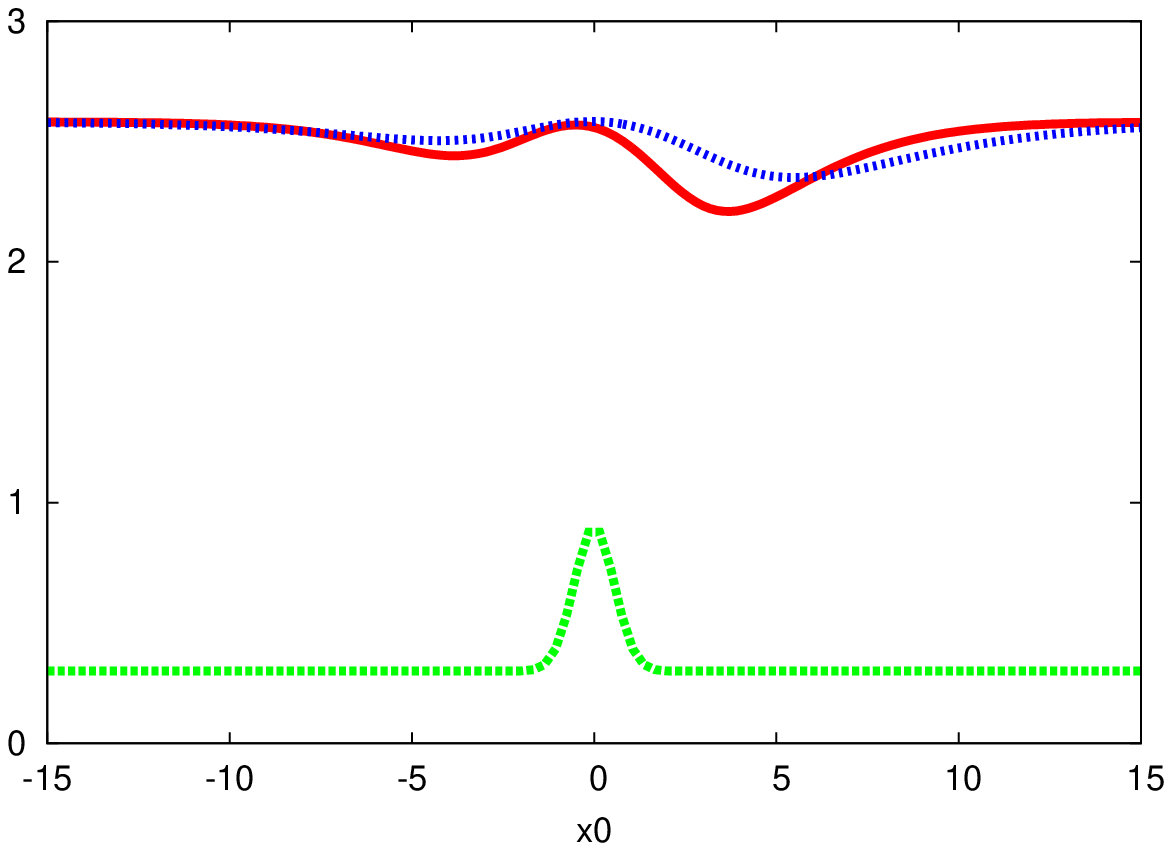,width=0.5\linewidth,angle=0}
\epsfig{file=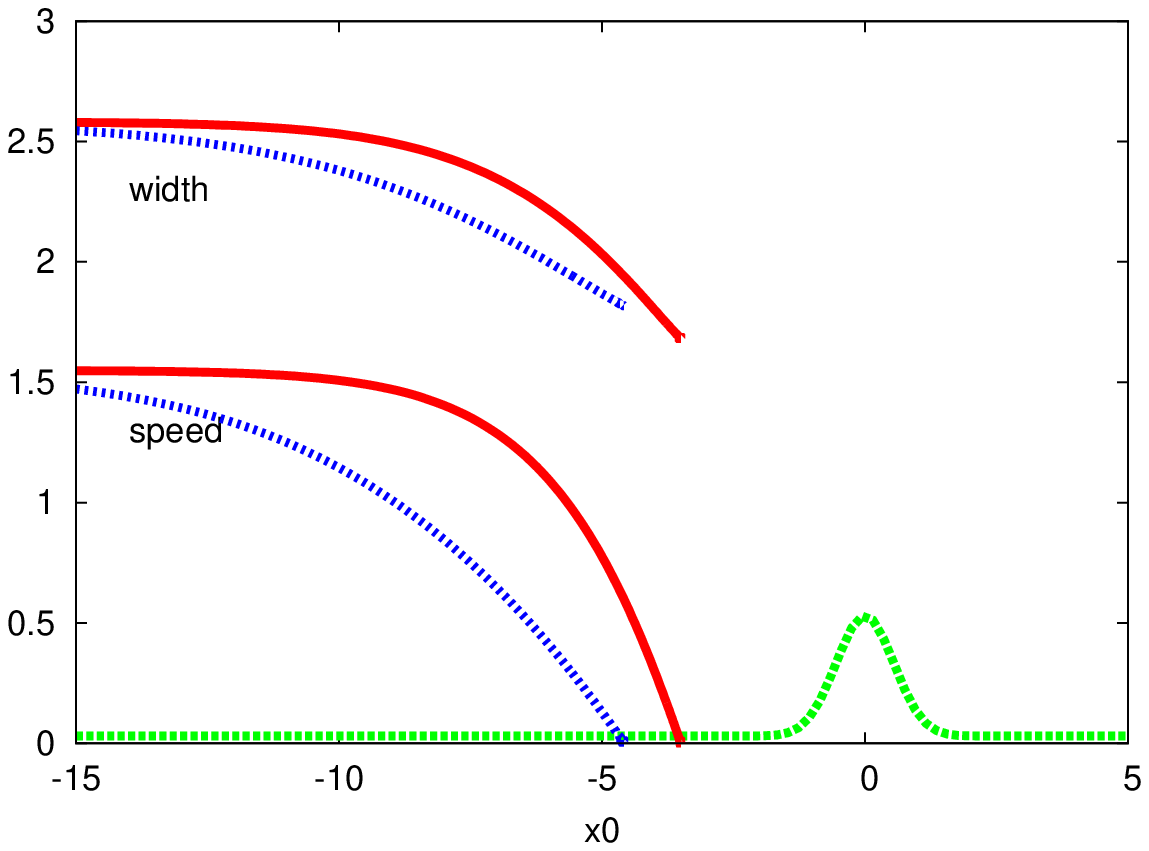,width=0.5\linewidth,angle=0}
}
\caption{Plots $(x_0,w)$ for the partial differential
equation \eqref{zeldo_s} in continuous line (red online)
and the collective variable equations \eqref{ode_dirac}
in dashed line (blue online) for a narrow gaussian defect shown at
the bottom of the plots. The left panel corresponds to $s_1=0,6$ and
$d=0,3$. The right panel corresponds to $s_1=5$. There the speed
$x_0'$ is also reported.}
\label{f10}
\end{figure}

For such narrow defects, the collective variable equations 
\eqref{ode_dirac} are less accurate than for a wide defect.
They do provide however the qualitative behavior, in particular the pinning of
the front.
Fig. \ref{f10} shows the plots for a gaussian defect drawn at the bottom.
The value of $s$ at infinity is $s_0=0,3$.
The left panel corresponds to a small amplitude $s_1=0,6$.
The right panel is for a much larger amplitude $s_1=5$ causing the pinning of the front.
For this particular plot we show the speed $x_0'$ computed using a centered
difference for both the partial differential equation and the collective variables.
They are multiplied by $10$ in the plot for clarity.
The defect has been divided by $10$.
The partial differential equation and the collective variables agree well for the width $w$ as a function of $x_0$.
The speed $x_0'$ is not so well approximated but it goes to zero for a pinning position $x_0$.

\begin{figure}[H]
\psfrag{x0}[r][r]{$x_0$}
\centerline{
\epsfig{file=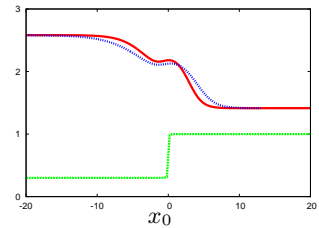,height=5cm,width=8cm,angle=0}
}
\caption{Plot $(x_0,w)$ for the partial differential
equation \eqref{zeldo_s} in continuous line (red online)
and the collective variable equations \eqref{ode_heaviside}
in dashed line (blue online) for a narrow "tanh" defect shown at
the bottom of the plot. The parameters are
$s_l=0,3,~ s_r=1$ -corresponding to a jump of $0,7$- and $d=0,1$.}
\label{f11}
\end{figure}

\begin{figure}[H]
\psfrag{x0}[r][r]{$x_0$}
\centerline{
\epsfig{file=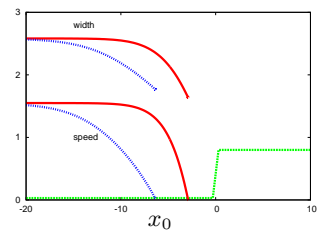,height=5cm,width=8cm,angle=0}
}
\caption{Plot $(x_0,w)$ for the partial differential
equation \eqref{zeldo_s} in continuous line (red online)
and the collective variable equations \eqref{ode_heaviside}
in dashed line (blue online) for a narrow "tanh" defect shown at
the bottom of the plot. The speed $x_0'$ is also reported
using the same color scheme.}
\label{f12}
\end{figure}

For narrow "tanh" defects, we reduce the collective variable equations to the ones for a Heaviside defect \eqref{ode_heaviside} as shown above.
The agreement between the curves $(x_0,w)$ for the partial differential equation \eqref{zeldo_s} and the collective variable equations \eqref{ode_heaviside} is good
as shown in Fig. \ref{f11}.
As we have seen in section \ref{numerical_analysis} the front can get pinned when
$s_1$ is large enough and this is predicted also by the 
the collective variable model. Such an example is shown in
Fig. \ref{f12} for a large and narrow defect where
$s_r=8,~d=0,1$ 
shown scaled by 0.1  at the bottom of the plot.
As previously the speed $x_0'$ estimated using finite differences has been
plotted in the graph to show pinning. 

\section{\label{critere}Estimates for front pinning and defect topography}

We now illustrate how the reduced model \eqref{xtwt_gen} can be used. Its
main advantage is that the parameters appear explicitly and therefore
their influence can be understood.
A first direct application is a simple criterion for the front pinning.

From the equation for the Dirac delta function defect, we can obtain
a rough estimate of the strength $\beta$ of the defect necessary to stop
the front. The evolution of the front from \eqref{ode_dirac}
shows that the front can stop $x_0'=0,\:w'=0$ when
\begin{equation} \label{ode_dirac2}
\left\{
\begin{aligned}
&\alpha w \left(\frac{1-2a}{2} \right) + \beta R\left(U\left(\frac{-x_0}{w}\right)\right) =\: 0 \\
&\frac{1}{3w} -  \alpha \frac{w}{6}  + \beta \left(1-2 U\left(\frac{-x_0}{w}\right)\right) R\left(U\left(\frac{-x_0}{w}\right)\right) =\: 0
\end{aligned}
\right.
\end{equation}
Therefore the front can stop when 
the two terms on the right hand side of the first equation balance each other.
In other words we need that
$$ \alpha w \left(\frac{1-2a}{2} \right) \leqslant -\beta \min (R). $$
So there exist a threshold for the pinning :
\be
\alpha w \left(\frac{1-2a}{2} \right) = -\beta_c \min (R) \label{beta_c}.
\ee
The minimum of $R(U(z))$ as a function of $z$ and its argument $z_{min}$ can be computed exactly, 
\begin{align}
z_{min}  = {} & \ln{\left( \frac{1+r}{a} \right)}, \\
\min (R) = {} & R(U(z_{min}))= - \frac{a^2(1+r)(a+r)}{(1+a+r)^3},
\end{align}
where $r=\sqrt{1-a+a^2}$.
Plugging $\beta=\beta_c$ into the second equation of \eqref{ode_dirac2} and assuming $\frac{-x_0}{w}=z_{min}$,
we get an expression for the width at pinning $w$. Using this we also get the front stopping position $x_0$.
For the critical $\beta$, the position and width at pinning are given by 
\begin{equation}
\left\{
\begin{aligned}
x_0 &= -\sqrt{\frac{2}{\alpha}} \frac {\ln{\left( \frac{1+r}{a} \right) }}{\sqrt{1+3(1-2a)\left( \frac{1-a+r}{1+a+r} \right)}}, \\
w   &= \sqrt{\frac{2}{\alpha}} \frac{1}{\sqrt{1+3(1-2a)\left( \frac{1-a+r}{1+a+r} \right)}}
\end{aligned}
\right.
\end{equation}
The critical $\beta$ is obtained from \eqref{beta_c} with the above value of $w$. It reads
\be
\beta \geqslant \beta_c = \sqrt{\frac{\alpha}{2}} \frac{(1-2a)(1+a+r)^3}{a^2(1+r)(a+r)\sqrt{1+3(1-2a)\left( \frac{1-a+r}{1+a+r} \right)}}.
\ee
These values are in good agreement with the values obtained numerically in the previous section, see Fig. \ref{f3}.
Precisely, with $a=0,3$, and $\alpha=0,3$ too, we find $\beta_c\approx5,88$, and the parameters at pinning $x_0\approx-3,47,\:w\approx1,89$.
Numerically, we found $\beta_c\approx6,18,\:x_0\approx-2,73,\:w\approx1,80$. This $\beta_c$ corresponds to $s_1\approx7,80$ because we set the width of the defect $d=0,1$.
Such a criterion can also be infered in the case of a Heaviside defect.
A pinning of the front is also possible with $\beta<0$ (as long as $\beta>-\alpha$ so that $s(x)$ remains positive).

\begin{figure}[H]
\psfrag{x}[r][r]{$x$}
\psfrag{s(x)}[r][r]{$s(x)$}
\centerline{
\epsfig{file=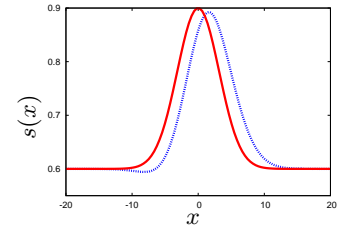,height=5cm,width=8cm,angle=0}
}
\caption{Defect topography $s(x)$ estimation from the solution of
the partial differential equation. The estimated defect topography
$s(x)$ using \eqref{sdx0w} is shown in dashed line (blue online)
while the "real" $s(x)$ is shown in continuous line (red online).}
\label{f_inv_pb}
\end{figure}

Another direct application of the collective variable model is that we can obtain the defect topography $s(x)$ from the observation of the front position $x_0$ and
width $w$.
We illustrate this on the example of a wide gaussian defect of the form \eqref{gaus_defect} with \eqref{zeldo_s} and the parameters $s_0=0,6,~s_1=0,3,~d=10$.
The position and width of the front are estimated using the least square fit on the solution of the partial differential equation \eqref{zeldo_s}.
From the collective variable equations in the adiabatic case \eqref{ode_adia} we get
\be\label{sdx0w}
s(x_0) = \frac{2}{1-2a} \frac{x_0'}{w}.
\ee
Using a centered difference approximation for the time derivative, we
obtain an estimate of $s(x_0)$. This estimate is compared to the "real"
$s(x)$
in Fig. \ref{f_inv_pb}.
As can be seen the agreement is very good. This approach can then be extended to other types of defects for physical or biological applications.

These two examples illustrate the power of these reduced models. The role
of the parameters is very easy to understand. One can easily solve the
inverse problem of estimating these parameters from measurements. Another
extension of this could be to control the front using these reduced
equations.

\section{\label{conclusion}Conclusion}

We have analyzed numerically the interaction of Zeldovich
reaction-diffusion front with a localized defect. The stability of the
front for different types of defects suggested that it has the
form of a generalized traveling wave with a time dependant
position and width. Using conservation
laws we obtained ordinary differential equations for these collective
variables. We further reduced these models for the cases of an adiabatic
defector a sharp "gaussian" or "tanh" defect. For these three cases the
position and width obtained by fitting the numerical solution agree
very well with the solutions of the collective variable equations.
Finally we illustrated how these reduced models can be used to
predict the pinning of the front on a large defect or to estimate the
defect topography from a time-series of front positions and width.

\appendix
\section{\label{numerical_details}Numerical procedures} 

The basis of the method is to discretize the spatial part of the
operator and keep the temporal part as such. We thereby transform
the partial differential equation into a system of ordinary
differential equations. This method allows to increase the
precision of the approximation in time and space independently and
easily.
We choose as space discretisation the finite volume
approximation where the operator is integrated over reference volumes.
The value of the function is assumed constant in each volume.
As solver for the system of differential equations, we use
the Runge-Kutta method of order 4-5 introduced by Dormand and Prince
which enables to control the local error by varying the time-step.
This has been implemented as the Fortran code DOPRI5 by Hairer and Norsett \cite{hairer}.

We introduce reference volumes $V_k$ (interval) whose centers are at
discretisation points $x_k = x_{min} + {h/2} + (k-1)h$, $1\leqslant k \leqslant n$
with $h = (x_{max}-x_{min})/n$
$$V_k=\left ]\frac{x_k+x_{k-1}}{2},\frac{x_{k+1}+x_k}{2}
\right[,~1 \leqslant k \leqslant n . $$
For a fixed t, we assume $u(x,t)$ to be constant on each volume $V_k$,
$u(x_k,t) = u_k$.
Integrating over $V_k$, we obtain for $1 < k< n$
\be \label{dis_rd_bulk} {\dot u_k} =\frac{u_{k+1} + u_{k-1} -2 u_{k}}{h^2}
+ s_k R(u_k) ,\ee
The boundary points $k=1,n$ are such that one satisfies the homogeneous
Neumann boundary conditions at $x=x_{min},x_{max}$.
We have 
\be {\dot u_1} = \frac{u_{2}-u_1}{h^2} + R(u_1)s_1 ,\ee
\be {\dot u_n}= \frac{-u_n+u_{n-1}}{h^2} + R(u_n)s_n.\ee

The fit of the solutions using the front profile is done using a 
least square method. For each time $t$ we introduce an error function
\be\label{energy_ls}
E(x_0,w)= \frac{1}{n} \sum_{i=1}^n (u_i(t)-u_k(x_i,t))^2
\ee
where the $u_i$ are the values of $u$ calculated numerically 
and 
$$u_k(x,t)=\frac{1}{1+\exp\left(\frac{x-x_0(t)}{w(t)}\right)}$$ 
is the exact kink solution. 
The expression $E$ is minimized using the Polak-Ribiere combination
of line minimisations \cite{numrec}. 
The fit of the solutions is done on the region such that $ 0.01 < u_k< 0.99$. 
The initial guesses for $x_0$ and $w$ are estimated from $u\approx 0.25$
and $u\approx 0.75$. The method converges in about 20 iterations and
the value of the energy is small, $ \min \: E \leqslant 10^{-4}$. Therefore
the fit is good.

\section{\label{analytical}Derivation of the collective variable equations}

The traveling wave is
\be
u(x,t) = U\left(\frac{x-x_0(t)}{w(t)}\right) \equiv U(z),
\ee
so that we have the relations for the partial derivatives
\begin{align}
u_t &= -\frac{1}{w} \left(x_0' + \frac{w'(x-x_0)}{w} \right) U' \left(\frac{x-x_0}{w} \right) ,\\
u_x &= \frac{1}{w} U' \left(\frac{x-x_0}{w} \right) ,\\
u_{xx} &= \frac{1}{w^2} U'' \left(\frac{x-x_0}{w} \right).
\end{align}
The first equation is obtained by integrating \eqref{rds} with respect to $x$
\begin{equation*}
\int_{-\infty}^{+\infty} u_t dx = [u_x]_{x=-\infty}^{+\infty} + \int_{-\infty}^{+\infty} s(x) R(u) dx
\end{equation*}
The term $[u_x]_{x=-\infty}^{+\infty}$ is zero because the front is flat away from the defect.
Introducing the partial derivatives we get
\begin{multline}
\int_{-\infty}^{+\infty} -\frac{x_0'}{w} U'\left(\frac{x-x_0}{w} \right) dx
- \int_{-\infty}^{+\infty} \frac{w'(x-x_0)}{w^2} U'\left(\frac{x-x_0}{w} \right) dx
\\
= \int_{-\infty}^{+\infty} s(x) R\left(U\left(\frac{x-x_0}{w} \right)\right) dx
\end{multline}
We compute the integrals by making the change of variables $x = w\:z + x_0$
\begin{multline*}
- x_0' \int_{-\infty}^{+\infty}  U'(z) dz
- w' \int_{-\infty}^{+\infty}  U'(z) z dz
\\
= w \int_{-\infty}^{+\infty} s(wz+x_0) R(U(z)) dz
\end{multline*}
Assuming that the front goes from 1 to 0 and that $U'(z)$ is even we get the final result
\begin{multline}\label{xt_gen}
x_0'  \int_{-\infty}^{+\infty}  U'(z) dz
+ w' \int_{-\infty}^{+\infty}  U'(z) z dz
\\
+ w \int_{-\infty}^{+\infty} s(wz+x_0) R(U(z)) dz = 0
\end{multline}

The second conservation law \eqref{urds} will yield the evolution of $w$. Proceeding as above we get
\begin{multline}\label{wt_gen}
x_0' \int_{-\infty}^{+\infty} U(z)U'(z) dz
+ w'  \int_{-\infty}^{+\infty} U(z)U'(z) z dz \\
+ w \int_{-\infty}^{+\infty} s(wz+x_0) U(z) R(U(z)) dz
= \frac{1}{w} \int_{-\infty}^{+\infty} U'^2(z) dz
\end{multline}
Note that we only assumed $U=1$ at $-\infty$ and $U=0$ at $\infty$ and that $U'$ is even. These assumptions are very general. In particular we have made no restrictions on the reaction term $R$.

For the Zeldovich reaction term $R(u) = u(1-u)(u-a)$ it 
is natural to assume that
$$U(z)=\frac{1}{1+\exp z}.$$
Then the integrals not involving $s$ can be computed and we obtain 
the final result.
\begin{equation}
\left\{
\begin{aligned}
x_0' &= w \int_{-\infty}^{+\infty} s(wz+x_0)R(U(z)) dz \\
w'   &= \frac{1}{3w} + w  \int_{-\infty}^{+\infty} s(wz+x_0) (1-2U(z))R(U(z)) dz
\end{aligned} 
\right.
\end{equation}

\end{document}